# Superconducting Quantum Interference Devices Made of Sb-doped Bi$_2$Se$_3$ Topological Insulator Nanoribbons

Nam-Hee Kim[1], Hong-Seok Kim[1*], Yasen Hou[2], Dong Yu[2], Yong-Joo Doh[1*]

[1]Department of Physics and Photon Science, Gwangju Institute of Science and Technology (GIST), Gwangju 61005, Korea

[2]Department of Physics, University of California, Davis, CA 95616, U.S.A.

[*]E-mail: canvas10@gist.ac.kr (H.-S. Kim), yjdoh@gist.ac.kr (Y.-J. Doh)

**Abstract**

We report the fabrication and characterization of superconducting quantum interference devices (SQUIDs) made of Sb-doped Bi$_2$Se$_3$ topological insulator (TI) nanoribbon (NR) contacted with PbIn superconducting electrodes. When an external magnetic field was applied along the NR axis, the TI NR exhibited periodic magneto-conductance oscillations, the so-called Aharonov-Bohm oscillations, owing to one-dimensional subbands. Below the superconducting transition temperature of PbIn electrodes, we observed supercurrent flow through TI NR-based SQUID. The critical current periodically modulates with a magnetic field perpendicular to the SQUID loop, revealing that the periodicity corresponds to the superconducting flux quantum. Our experimental observations can be useful to explore Majorana bound states (MBS) in TI NR, promising for developing topological quantum information devices.

## 1. Introduction

Topological insulators (TIs) are bulk insulators having metallic surface states that are topologically protected by time-reversal symmetry [1]. The topological surface states (TSSs) are not subjected to backscattering by nonmagnetic impurities because the spin of the surface electrons are perpendicularly aligned to their translational momentum [2]. This spin-momentum locking property leads to a highly coherent quantum transport, making TIs promising candidates for quantum information devices [3]. For TI nanoribbon (NR) structure, the TSSs are modified as the one-dimensional (1D) subbands (or 1D surface modes) owing to the quantum confinement effect along the circumference of the NR, combined with the Aharonov-Bohm (A-B) phase and Berry's phase of $\pi$ [4, 5]. Thus, under an applied external magnetic field along the NR axis, the magneto-conductance of TI NR oscillates with magnetic flux threading the cross-sectional area of the NR, the so-called A-B oscillations in TI NR. The A-B oscillations provide an evidence for the TSS in TI NR [6-9].

When TI materials are combined with conventional *s*-wave superconductors, they are expected to form a topological superconducting state hosting the Majorana bound state (MBS) [10], which is a building block for developing topological quantum information technology [11]. After the first observations of supercurrent through TI materials [12, 13], nonlocal Fraunhofer-like pattern [14] and macroscopic quantum tunneling behavior [15] have been observed in TI-based Josephson junctions (JJs). As an evidence for the MBS, abnormal Shapiro steps due to $4\pi$-periodic current-phase relation (CPR) have been reported in TI-based JJs [16, 17]. It would be more plausible to directly measure such CPR using a dc superconducting quantum interference device (SQUID) consisting of two parallel JJs [18]. So far, TI-based SQUID has been studied using micro flakes or thin film of TI [19-22]. TI NR with reduced dimensions, however, is more promising for exploring MBS in TI, which results from an enhanced surface-to-volume ratio in NR.

There are three requisites for successful development of topological SQUID based on TI NR. Firstly, TSS should be guaranteed to exist in TI NR [23]. To confirm the existence of the 1D helical surface modes in TI NR, it is necessary to demonstrate A-B oscillations with the axial magnetic field in the normal state. Secondly, highly transparent contact should be formed at the interface between TI NR and the superconducting electrode to induce strong superconducting proximity effect, which is essential to form MBS. Lastly, the superconductivity of the superconducting electrode should be maintained until the magnetic flux through the cross-sectional area of TI NR reaches half flux quantum, $\Phi = h/2e$, where $h$ is the Planck's constant, and $e$ is the electric charge. This criterion is required to form a gapless helical 1D mode in TI NR [4].

In this work, we present an experimental study of hybrid SQUIDs made of Sb-doped $Bi_2Se_3$ TI NR and PbIn alloy superconductor. Our TI NR devices exhibited A-B oscillations with magnetic field along the NR axis in the normal state, which is attributed to the formation of 1D helical surface states in TI NR. Under an external magnetic field perpendicularly applied to the SQUID loop, periodic oscillation of the critical current was observed with a magnetic-flux periodicity of $h/2e$ in the loop. At high magnetic fields, the Fraunhofer pattern was superimposed on the SQUID oscillations owing to critical current modulations of individual JJs consisting of the SQUID. These results indicate that the TI NR-based SQUID can provide a promising new platform for exploring MBS and developing topological quantum computation technology.

## 2. Experiments

Sb-doped $Bi_2Se_3$ NRs [24] were synthesized via a chemical vapor deposition method in a horizontal tube furnace (Lindberg/ Blue M). Mixed $Bi_2Se_3$ powder (100 mg, 99.999 %, Alfa Aesar) and Sb powder (5-30 mg, 99.999 %, Johnson Matthey Inc.) were placed at the center of the tube furnace. Se pellets (200 mg, 99.999 %, Johnson Matthey Inc.) were placed upstream at a distance of 16 cm. A Si substrate with a 10 nm-thick gold film layer deposited by electron beam evaporation was placed 7 cm downstream as the growth substrate. The tube was initially pumped down to a base pressure of about 40 mTorr and then filled with Ar gas (99.995 %) to ambient pressure. Under the flow of Ar gas at 150 sccm, the furnace was heated to 680 °C in 8 minutes and was maintained at the peak temperature for 5 hours. After that, the furnace was cooled to room temperature over approximately 3 hours. The representative scanning electron microscopy (SEM) image of the Sb-doped $Bi_2Se_3$ NRs after the growth process is displayed in Fig. 1a. The inset showed that the NR had a rectangular cross section. The widths ($w$) of the NRs used for this experiment ranged from 370 to 500 nm and the thicknesses ($t$) from 120 to 230 nm (see Table 1). The overall length of the TI NR was up to tens of micrometers. Energy-dispersive X-ray (EDX) spectroscopy for the NRs revealed that the atomic percentages (at. %) for Bi, Sb, and Se were 35.2 ± 1.6, 6.5 ± 1.3, 58.3 ± 1.1, respectively, indicating the NR composition as $(Bi_{0.84}Sb_{0.16})_2Se_3$ (not shown here).

After the growth of Sb-doped $Bi_2Se_3$ NRs, individual NRs were mechanically transferred to a highly $n$-doped silicon substrate covered by a 290 nm-thick $SiO_2$ layer with pre-patterned Ti/Au bonding pads. Source and drain electrodes were patterned by standard electron-beam lithography followed by electron-beam evaporation of 300 nm-thick PbIn [25]. A 10 nm-thick Au film was used as a capping layer to avoid oxidation. The PbIn alloy source was made using a mixture of Pb and In pellets with a weight ratio of 1:1. Before the metal deposition, the NR surface was cleaned by oxygen plasma treatment to remove the electron-beam resist residues that could remain. Then, the NR was treated using a 6:1 buffered oxide etch (BOE) for 7 s to remove a native oxide layer that was assumed to be present. Typical SEM image of TI NR-based SQUID after completion of the device fabrication process is shown in Fig. 1b. The PbIn electrodes showed a superconducting transition at temperature $T_c$ = 6.8 K, while the critical magnetic field was estimated to be $H_{c,perp}$ = 0.74 T ($H_{c,para}$ = 1.2 T) under a perpendicular (parallel) magnetic field [26]. The electrical transport properties of the TI NR-based SQUIDs were measured using a closed-cycle $^4$He cryostat (Seongwoo Instruments Inc.) and $^3$He refrigerator (Cryogenic Ltd.) which have a base temperature of 2.4 and 0.3 K, respectively.

For low-noise measurement, low-pass RC and π filters were connected in series with the measurement leads [27]. The device geometry and characteristic parameters are summarized in Table 1.

### 3. Results and discussion

When the magnetic field is perpendicularly applied to the surface of TI, the breaking of time-reversal symmetry results in negative magneto-conductance (MC) near the zero magnetic field, which is interpreted in the framework of the weak anti-localization (WAL) effect [28]. Since the WAL effect is due to a two-dimensional surface state, only the normal component of magnetic ($B$) field, $B\sin\theta$, contributes to the negative MC behavior, where $\theta$ is the angle of the magnetic field relative to the substrate. Thus, the MC curves obtained at different angles were merged into a single one when those were replotted against $B\sin\theta$ [9, 29]. Figure 1c shows the MC curves, $\Delta\sigma = \sigma(B) - \sigma(0)$, of sample **D1** contacted with Ti(10 nm)/Au(250 nm) electrodes. The width and thickness of the NR were $w = 607$ nm and $t = 90$ nm, respectively, while the channel length was $L = 4.89$ μm. All MC curves observed at different $\theta$'s exhibited a cusp-like feature near zero magnetic field and merged into a single one when those were plotted with $B\sin\theta$, indicating the surface effect was dominant over the negative MC behavior.

According to the Hikami–Larkin–Nagaoka (HLN) theory [28], the WAL effect in two-dimensional (2D) limit can be expressed by

$$\Delta\sigma_{2D} = \alpha \frac{e^2}{2\pi^2 \hbar} \left[ \ln\left(\frac{\hbar}{4eB_n L_\varphi^2}\right) - \Psi\left(\frac{1}{2} + \frac{\hbar}{4eB_n L_\varphi^2}\right) \right], \tag{1}$$

where $\alpha$ is the dimensionless parameter corresponding to 0.5 for a single conducting channel, $\hbar$ is the reduced Planck's constant, $B_n$ is the normal component of the magnetic field, $\Psi$ is the digamma function, and $L_\varphi$ is the phase coherence length. Our experimental MC data were fitted well by the

Eq. (1) with $\alpha = 1.35$ and $L_\varphi = 447$ nm at $T = 3.1$ K (see solid line in Fig. 1c). The obtained value of $L_\varphi$, which was similar to those previously reported using $Bi_2Se_3$ [30, 31], satisfied the 2D limit condition ($w > L_\varphi$). The value of $\alpha$ exceeded the expected value of 1, which is due to the two decoupled conducting channels at the top and bottom surfaces of TI sample. This difference can be attributed to an additional Rashba-split trivial 2D channel [32] or coherent transport in the bulk [30].

When we applied the axial magnetic field, $B_{axial}$, along the NR axis of device **D2**, the magneto-resistance (MR) oscillated with a period of $\Delta B_{axial} = 0.17 \pm 0.01$ T, as shown in Fig. 1d. As the dimensions of the NR were given by $w = 400$ nm and $t = 79$ nm, the magnetic-flux period corresponded to $\Phi_{period} = wt \times \Delta B_{axial} = 1.3$ $h/e$. This $\Phi_{period}$ value was close to the magnetic flux quantum in the normal state, $\Phi_0 = h/e$, as expected in the A-B oscillation theory of TI NR [4, 5]. The difference can be explained by the native oxide layer on the surface of TI NR, where the layer thickness is estimated to be about 5 nm [8, 33]. Then, we obtained $\Phi_{period} = 1.1$ $h/e$, which was considerably closer to the theoretical expectation. Our observation of the A-B oscillations confirmed the existence of 1D surface modes [4] in our Sb-doped $Bi_2Se_3$ NR.

Schematic and dimensions of TI-NR SQUID's are shown in the inset of Fig. 2a and Table. 1, respectively. Two JJs, fabricated on top of the Sb-doped $Bi_2Se_3$ TI NR with PbIn superconducting electrodes, were connected to each other to form a superconducting loop [25, 26]. Geometrical dimensions of our devices are listed in Table 1. Figure 2a shows the current–voltage ($I$–$V$) characteristic curves of device **Sq2** measured at $T = 2.4$ K. The maximum supercurrent, so-called the critical current, was obtained as $I_C = 0.40$ µA in the absence of magnetic field. When the magnetic field was perpendicularly applied to the SQUID loop, $I_C$ gradually decreased and disappeared when the magnetic flux through the loop reached $\Phi = 0.5$ $\Phi_S$, where $\Phi_S = h/2e$ is the superconducting flux quantum. For $0.5 < \Phi/\Phi_S < 1$, $I_C$ increased and returned to its original value with $I_C(\Phi_S) = I_C(0)$. Figure 2b, a color plot of differential resistance d$V$/d$I$, shows the periodic modulation of $I_C$ as a function of $B$ and $\Phi$. In this figure, the dark regions corresponds to the superconducting state with $dV/dI = 0$ and oscillates with the periodicity of $\Phi_S$ and $\Delta B = 4.7$ Oe, which is a characteristic feature of the conventional SQUID [18]. The effective area for the magnetic flux was estimated to be $A_{\text{eff}} = 4.4$ μm$^2$ using a relation of $\Phi_S = \Delta B \times A_{\text{eff}}$, while the inner geometric area of the SQUID loop was $A = 1.71$ μm$^2$. This difference can be explained by the London penetration depth ($\lambda_L$), where the magnetic field penetrates the superconducting electrodes. We estimated $\lambda_L = 200$ nm for PbIn electrode from $A_{\text{eff}} = (L_S + 2\lambda_L) \times (W_S + 2\lambda_L)$, which was similar to the values obtained from previous work [34].

The periodic modulation of $I_C(\Phi)$ can be described by [35]

$$I_C(\Phi) = \left[ \left( I_{C1} - I_{C2} \right)^2 + 4 I_{C1} I_{C2} \cos^2 \left( \pi \Phi / \Phi_S \right) \right]^{1/2} \tag{2}$$

where $I_{C1}$ and $I_{C2}$ are the critical currents of two JJs consisting of TI-NR SQUID. The fitting results are shown in Fig. 2b (solid white lines) with parameters of $I_{C1} = 204$ nA and $I_{C2} = 198$ nA. From this, we can deduce that the device **Sq2** belongs to a symmetric case with $I_{C2}/I_{C1} = 0.97$. The self-inductance of the SQUID loop was estimated as $L_S \sim 3.5$ pH from the loop geometry, indicating that the screening parameter $\beta_L = 2\pi L_S I_0 / \Phi_S = 2.2 \times 10^{-3} \ll 1$, where $I_0 = (I_{C1} + I_{C2})/2$ is the average critical current. Thus, the self-inductance effect was negligible in our experiment.

The $I_C(\Phi)$ modulation can be reflected by voltage ($V$) modulation under a dc current bias $I_{dc}$, as shown in Fig. 2c. As the SQUID acts as a flux-to-voltage transducer, the output voltage $V$ oscillates with the magnetic flux through the SQUID loop with a period of $\Phi_S = h/2e$ [18]. The oscillation amplitude increased with $I_{dc}$ and became maximum near $I_C$. The maximum sensitivity of the TI-NR SQUID was obtained as $\left| \partial V / \partial \Phi \right| = 17$ μV/$\Phi_S$ (see Fig. 2d), which is comparable to previous studies of dc SQUID based on Bi$_2$Te$_3$ flake [19] or PbS nanowire [25].

Figure 3a displays the *I-V* curves obtained from the **Sq4** device at different magnetic fields and lower temperature, $T = 0.3$ K. The maximum $I_C$ was given by $I_{C,max} = I_C (\Phi = 0) = 7.2$ μA in the absence of magnetic field, which was an order of magnitude larger than that of **Sq2**. $I_C$ monotonously decreased with increasing $\Phi$ from 0 to $0.5\ \Phi_S$. However, it was noted that a finite $I_C$ remained even at $\Phi = 0.5\ \Phi_S$, resulting in the absence of the supercurrent OFF state. The color plot of d$V$/d$I$ *vs*. $\Phi$, as shown in Fig. 3b, clearly displays the $I_C (\Phi)$ modulation and the existence of non-zero supercurrent at $\Phi = (n + 1/2)\ \Phi_S$, where $n$ is an integer. The residual supercurrent, $I_{C,min}$, can be attributed to a self-inductance screening effect or asymmetric JJ's in SQUID [35]. As the screening parameter was quite small ($\beta_L \ll 1$), the self-inductance effect was negligible in this experiment. Two different $I_C$'s for each JJ can be estimated from $I_{C1} = (I_{C,max} + I_{C,min})/2$ and $I_{C2} = (I_{C,max} - I_{C,min})/2$, respectively [35]. Thus, we obtained $I_{C1} = 4.8$ μA and $I_{C2} = 2.4$ μA, respectively, which means $I_{C2}/I_{C1} = 0.50$. The **Sq4** device belongs to asymmetric dc-SQUID regime in contrast to **Sq2**. When plotting the Eq. (2) using the above estimates of $I_{C1}$ and $I_{C2}$ (see the solid white lines), it fitted well the $I_C (\Phi)$ modulation behavior in Fig. 3b.

The $V (\Phi)$ modulation of **Sq4** is shown in Fig. 3c with varying $I_{dc}$. The modulation periodicity was $\Phi_S = h/2e$ in terms of magnetic flux, while its amplitude was maximized near $I_{dc} \approx I_C$. Figure 3d depicts the SQUID sensitivity as a function of $I_{dc}$. The maximum value of $|\partial V/\partial \Phi|$ was about 152 μV/$\Phi_0$, which is almost an order of magnitude larger than those of **Sq2** and in Ref.[19]. This difference can be explained by a larger $I_C R_N \sim 39.0$ μV value of device **Sq4** in comparison with **Sq2**, where $R_N$ means the normal-state resistance of SQUID. It should be noted that the maximum $I_C R_N$ value obtained from our TI-NR SQUID was still larger than that of TI flake-based SQUID's [21, 36], which is attributed to a relatively large superconducting gap of PbIn and formation of highly transparent contacts between TI NR and superconducting electrodes. Hence, we suggest that the superconducting proximity effect through TI NR would be more advantageous to explore MBS in TI.

When we applied higher magnetic field to **Sq2**, a beat-like interference pattern was observed in the $V(\Phi)$ modulation, as shown in Fig. 4a, where an envelope was superposed on the previous SQUID oscillations. The fast Fourier transform (FFT) analysis of the $V(\Phi)$ modulation (see the inset of Fig. 4a) revealed that there were two different frequencies centered at $\Delta B^{-1} = 0.011$ and $0.22$ Oe$^{-1}$, corresponding to the periods of $\Delta B = 91$ and $4.6$ Oe, respectively. Another device (**Sq4**) also showed similar behavior (not shown here). The latter mode with short periodicity was responsible for the SQUID oscillations in Fig. 2c, while the former one was attributed to the magnetic field-dependent $I_C$ modulation of individual JJ's, the so-called Fraunhofer pattern. It is well known that $I_C$ of JJ becomes minimum when the magnetic flux through the junction's lateral area reaches an integer multiple of magnetic flux quantum, i.e., $\Phi = n\,\Phi_S$ with an integer $n$ [18]. Then the field interval between two successive peaks in the envelope, $\Delta B^* = 82$ Oe, corresponds to a single flux quantum through the JJ in the SQUID. As a result, we have a relation of $\Phi_S = \Delta B^*(L_{JJ} + 2\lambda_L)w$, where $L_{JJ} = (L_1 + L_2)/2$ is the average junction length of two JJ's and $w$ is the width of TI NR, resulting in $\lambda_L = 203$ nm, which is consistent with previous result obtained from the low-field SQUID oscillations.

When we combine the Fraunhofer-type $I_C$ modulation with the SQUID oscillation, Eq. (2) can be modified as follows [35]:

$$I_C(\Phi) = \left[(I_{C1} - I_{C2})^2 + 4 I_{C1} I_{C2} \cos^2(\pi\Phi/\Phi_S)\right]^{1/2} \left| \sin\left(\frac{\pi\Phi_J}{\Phi_S}\right) \Big/ \left(\frac{\pi\Phi_J}{\Phi_S}\right) \right| \qquad (3)$$

where $\Phi_J = (L_{JJ} + 2\lambda_L)wB$ is the magnetic flux through the single JJ. This new equation fits well the $I_C(B)$ modulation data obtained in the wider range of $B$ field, as can be seen in Fig. 4b. Thus, the CPR study of TI-NR SQUID should consider the SQUID oscillations together with the $I_C$ modulation effect of the individual junctions. Similar results can be found in other reports of TI-flake SQUID [20, 21].

In conclusion, we reported successful fabrication and characterization of SQUID's based on Sb-doped $Bi_2Se_3$ TI NR, contacted with PbIn superconducting electrodes. The angle-dependent WAL effect and the A-B oscillations confirm the existence of topological surface states in the TI NR. The SQUID oscillations with a periodicity of superconducting flux quantum, $h/2e$, have been investigated in terms of $I_C(B)$ and $V(B)$ modulations for both symmetric and asymmetric junction devices. Relatively large values of $I_CR_N$ product and the SQUID sensitivity indicates the formation of highly transparent superconducting contacts on TI NR. Our observations suggest that TI NR-based SQUID's are promising for exploring Majorana bound states in TI and developing topological quantum information devices.

## Acknowledgment

This work was supported by the NRF of Korea (Grant 2018R1A3B1052827), GIST-Caltech Collaboration Research grant (for Y.D.) and the U.S. National Science Foundation (Grant DMR-1710737 for D.Y.).

# Figure Captions

**Fig. 1.** (a) SEM image of the Sb-doped $Bi_2Se_3$ NR. Inset: Tilted-SEM image of the NR. Scale bar: 500 nm. (b) Representative SEM image of the SQUID based on Sb-doped $Bi_2Se_3$ NR and PbIn superconducting electrodes. (c) MC, $\Delta\sigma = \sigma(B) - \sigma(0)$, curves at various magnetic field orientations, as a function of perpendicular magnetic field component, $B\sin\theta$ (sample **D1**). The solid line is the best fit to Eq. (1). (d) MR curve under the axial magnetic field along the NR axis, $B_{axial}$, measured at $T = 2.3$ K (sample **D2**). The arrows indicate resistance dips occurring at $B_{axial} = 0.08, 0.24, 0.42$ T.

**Fig. 2.** (a) Current–voltage (*I–V*) curves for different magnetic fluxes through the TI-NR SQUID loop at $T = 2.4$ K (sample **Sq2**). Inset: Schematic of TI-NR SQUID. (b) Color plot of $dV/dI$ as a function of bias current and magnetic field (or $\Phi/\Phi_S$). The dark region indicates the supercurrent regime and the white solid line is a calculation result using Eq. (2). (c) Voltage modulation as a function of $\Phi/\Phi_S$ under a dc current bias of $I_{dc} = 0.2$ μA (black), 0.4 μA (red), 0.6 μA (blue), 0.8 μA (green) and 1.0 μA (magenta). (d) Squid sensitivity as a function of $I_{dc}$. Error bars are indicated.

**Fig. 3.** (a) *I–V* curves of device **Sq4** at various magnetic fluxes and $T = 0.3$ K. (b) Color plot of $dV/dI$ as a function of $I_{dc}$, $B$, and $\Phi/\Phi_S$. The white line presents calculation results using Eq. (2) considering $I_C$ asymmetry. (c) Output voltage as a function of $\Phi/\Phi_S$ with $I_{dc} = 2.8$ μA (black), 4.9 μA (red), 7.0 μA (blue), 7.7 μA (green) and 8.4 μA (magenta), respectively. (d) SQUID sensitivity as a function of $I_{dc}$.

**Fig. 4.** (a) Voltage modulation of **Sq2** as a function of $B$ field at $T = 2.4$ K. Inset: FFT analysis of $V(B)$ modulation. (b) Magnetic field dependence of $I_C$ of **Sq2**. The solid line is a fitting result using Eq. (3) (see text).

# References

[1] M.Z. Hasan, C.L. Kane, Colloquium: topological insulators, Reviews of Modern Physics, 82 (2010) 3045.

[2] P. Roushan, J. Seo, C.V. Parker, Y.S. Hor, D. Hsieh, D. Qian, A. Richardella, M.Z. Hasan, R.J. Cava, A. Yazdani, Topological surface states protected from backscattering by chiral spin texture, Nature, 460 (2009) 1106.

[3] C.W.J. Beenakker, Search for Majorana Fermions in Superconductors, Annual Review of Condensed Matter Physics, 4 (2013) 113-136.

[4] J.H. Bardarson, P.W. Brouwer, J.E. Moore, Aharonov-Bohm Oscillations in Disordered Topological Insulator Nanowires, Physical Review Letters, 105 (2010) 156803.

[5] R. Egger, A. Zazunov, A.L. Yeyati, Helical Luttinger Liquid in Topological Insulator Nanowires, Physical Review Letters, 105 (2010) 136403.

[6] S.S. Hong, Y. Zhang, J.J. Cha, X.L. Qi, Y. Cui, One-Dimensional Helical Transport in Topological Insulator Nanowire Interferometers, Nano Letters, 14 (2014) 2815-2821.

[7] L.A. Jauregui, M.T. Pettes, L.P. Rokhinson, L. Shi, Y.P. Chen, Magnetic field-induced helical mode and topological transitions in a topological insulator nanoribbon, Nature Nanotechnology, 11 (2016) 345-351.

[8] H.-S. Kim, H.S. Shin, J.S. Lee, C.W. Ahn, J.Y. Song, Y.-J. Doh, Quantum electrical transport properties of topological insulator $Bi_2Te_3$ nanowires, Current Applied Physics, 16 (2016) 51-56.

[9] J. Kim, A. Hwang, S.-H. Lee, S.H. Jhi, S. Lee, Y.C. Park, S.-I. Kim, H.-S. Kim, Y.-J. Doh, J. Kim, B. Kim, Quantum Electronic Transport of Topological Surface States in beta-$Ag_2Se$ Nanowire, Acs Nano, 10 (2016) 3936-3943.

[10] L. Fu, C.L. Kane, Superconducting proximity effect and Majorana fermions at the surface of a topological insulator, Phys. Rev. Lett., 100 (2008) 096407.

[11] J. Alicea, New directions in the pursuit of Majorana fermions in solid state systems, Reports on Progress in Physics, 75 (2012).

[12] M. Veldhorst, M. Snelder, M. Hoek, T. Gang, V.K. Guduru, X.L. Wang, U. Zeitler, W.G. van der Wiel, A.A. Golubov, H. Hilgenkamp, A. Brinkman, Josephson supercurrent through a topological insulator surface state, Nature Materials, 11 (2012) 417-421.

[13] J.R. Williams, A.J. Bestwick, P. Gallagher, S.S. Hong, Y. Cui, A.S. Bleich, J.G. Analytis, I.R. Fisher, D. Goldhaber-Gordon, Unconventional Josephson effect in hybrid superconductor-topological insulator devices, Phys. Rev. Lett., 109 (2012) 056803.

[14] J.H. Lee, G.-H. Lee, J. Park, J. Lee, S.-G. Nam, Y.-S. Shin, J.S. Kim, H.-J. Lee, Local and nonlocal Fraunhofer-like pattern from an edge-stepped topological surface Josephson current distribution, Nano Lett., 14 (2014) 5029-5034.

[15] J. Kim, B.-K. Kim, H.-S. Kim, A. Hwang, B. Kim, Y.-J. Doh, Macroscopic quantum tunneling in superconducting junctions of β-$Ag_2Se$ topological insulator nanowire, Nano Lett., 17 (2017) 6997-7002.

[16] J. Wiedenmann, E. Bocquillon, R.S. Deacon, S. Hartinger, O. Herrmann, T.M. Klapwijk, L. Maier, C. Ames, C. Brüne, C. Gould, A. Oiwa, K. Ishibashi, S. Tarucha, H. Buhmann, L.W. Molenkamp, 4π-periodic Josephson supercurrent in HgTe-based topological Josephson junctions, Nature Communications, 7 (2016) 10303.

[17] E. Bocquillon, R.S. Deacon, J. Wiedenmann, P. Leubner, T.M. Klapwijk, C. Brüne, K. Ishibashi, B. H., M.L. W., Gapless Andreev bound states in the quantum spin Hall insulator HgTe, Nature Nanotechnology, 12 (2016) 137-143.

[18] M. Tinkham, Introduction to Superconductivity, Dover Publications, 2004.

[19] M. Veldhorst, C. Molenaar, X. Wang, H. Hilgenkamp, A. Brinkman, Experimental realization of superconducting quantum interference devices with topological insulator junctions, Applied Physics Letters, 100 (2012) 072602.

[20] F. Qu, F. Yang, J. Shen, Y. Ding, J. Chen, Z. Ji, G. Liu, J. Fan, X. Jing, C. Yang, Strong superconducting proximity effect in Pb-$Bi_2Te_3$ hybrid structures, Scientific reports, 2 (2012) 339.

[21] C. Kurter, A. Finck, Y.S. Hor, D.J. Van Harlingen, Evidence for an anomalous current–phase relation in topological insulator Josephson junctions, Nature communications, 6 (2015) 7130.

[22] R. Klett, J. Schonle, A. Becker, D. Dyck, K. Borisov, K. Rott, D. Ramermann, B. Buker, J. Haskenhoff, J. Krieft, T. Hubner, O. Reimer, C. Shelkar, J.M. Schmalhorst, A. Huetten, C. Felser, W. Wernsdorfer, G. Reiss, Proximity-Induced Superconductivity and Quantum Interference in Topological Crystalline Insulator SnTe Thin-Film Devices, Nano Letters, 18 (2018) 1264-1268.

[23] A. Cook, M. Franz, Majorana fermions in a topological-insulator nanowire proximity-coupled to an s-wave superconductor, Physical Review B, 84 (2011) 201105.

[24] S.S. Hong, J.J. Cha, D. Kong, Y. Cui, Ultra-low carrier concentration and surface-dominant transport in antimony-doped $Bi_2Se_3$ topological insulator nanoribbons, Nature Communications, 3 (2012) 757.

[25] H.-S. Kim, B.-K. Kim, Y. Yang, X. Peng, S.-G. Lee, D. Yu, Y.-J. Doh, Gate-tunable superconducting quantum interference devices of PbS nanowires, Applied Physics Express, 9 (2016) 023102.

[26] N.-H. Kim, B.-K. Kim, H.-S. Kim, Y.-J. Doh, Fabrication and characterization of PbIn-Au-PbIn superconducting junctions, Progress in Superconductivity and Cryogenics, 18 (2016) 5-8.

[27] D. Jeong, J.-H. Choi, G.-H. Lee, S. Jo, Y.-J. Doh, H.-J. Lee, Observation of supercurrent in PbIn-graphene-PbIn Josephson junction, Physical Review B, 83 (2011) 094503.

[28] S. Hikami, A.I. Larkin, Y. Nagaoka, Spin-orbit interaction and magnetoresistance in the two dimensional random system, Progress of Theoretical Physics, 63 (1980) 707-710.

[29] J.J. Cha, D.S. Kong, S.S. Hong, J.G. Analytis, K.J. Lai, Y. Cui, Weak Antilocalization in $Bi_2(Se_xTe_{1-x})_3$ Nanoribbons and Nanoplates, Nano Letters, 12 (2012) 1107-1111.

[30] H. Steinberg, J.B. Laloe, V. Fatemi, J.S. Moodera, P. Jarillo-Herrero, Electrically tunable surface-to-bulk coherent coupling in topological insulator thin films, Physical Review B, 84 (2011) 233101.

[31] J.J. Cha, M. Claassen, D.S. Kong, S.S. Hong, K.J. Koski, X.L. Qi, Y. Cui, Effects of Magnetic Doping on Weak Antilocalization in Narrow $Bi_2Se_3$ Nanoribbons, Nano Letters, 12 (2012) 4355-4359.

[32] J. Lee, J. Park, J.-H. Lee, J.S. Kim, H.-J. Lee, Gate-tuned differentiation of surface-conducting states in $Bi_{1.5}Sb_{0.5}Te_{1.7}Se_{1.3}$ topological-insulator thin crystals, Physical Review B, 86 (2012) 245321.

[33] M.L. Tian, W. Ning, Z. Qu, H.F. Du, J. Wang, Y.H. Zhang, Dual evidence of surface Dirac states in thin cylindrical topological insulator $Bi_2Te_3$ nanowires, Scientific Reports, 3 (2013) 1212.

[34] B.-K. Kim, H.-S. Kim, Y.M. Yang, X.Y. Peng, D. Yu, Y.-J. Doh, Strong Superconducting Proximity Effects in PbS Semiconductor Nanowires, Acs Nano, 11 (2017) 221-226.

[35] A. Barone, G. Paterno, Physics and applications of the Josephson effect, Wiley New York, 1982.

[36] L. Galletti, S. Charpentier, M. Iavarone, P. Lucignano, D. Massarotti, R. Arpaia, Y. Suzuki, K. Kadowaki, T. Bauch, A. Tagliacozzo, Influence of topological edge states on the properties of $Al/Bi_2Se_3/Al$ hybrid Josephson devices, Physical Review B, 89 (2014) 134512.

**Table 1**. Physical parameters of the SQUIDs: $w$ ($t$) is the width (thickness) of TI NR. $L_1$ and $L_2$ are the channel length of each Josephson junction. $L_S$ and $W_S$ are the length and width of the SQUID, respectively. $T_{base}$ is the base temperature for $I_C$ measurement.

| Sample | $w$ (nm) | $t$ (nm) | $L_1$ (nm) | $L_2$ (nm) | $L_S$ (μm) | $W_S$ (μm) | $I_C$ (μA) | $I_C R_N$ (μV) | $T_{base}$ (K) |
|---|---|---|---|---|---|---|---|---|---|
| Sq1 | 506 | 200 | 157 | 121 | 1.34 | 3.43 | 0.45 | 3.2 | 2.4 |
| Sq2 | 495 | 200 | 114 | 89 | 1.36 | 1.26 | 0.40 | 4.6 | 2.4 |
| Sq3 | 414 | 235 | 290 | 257 | 1.59 | 1.41 | 2.60 | 14.8 | 0.3 |
| Sq4 | 368 | 235 | 243 | 292 | 1.46 | 1.66 | 7.20 | 39.0 | 0.3 |
| Sq5 | 391 | 121 | 260 | 260 | 1.30 | 1.70 | 0.10 | 1.6 | 2.5 |

**Fig. 1**

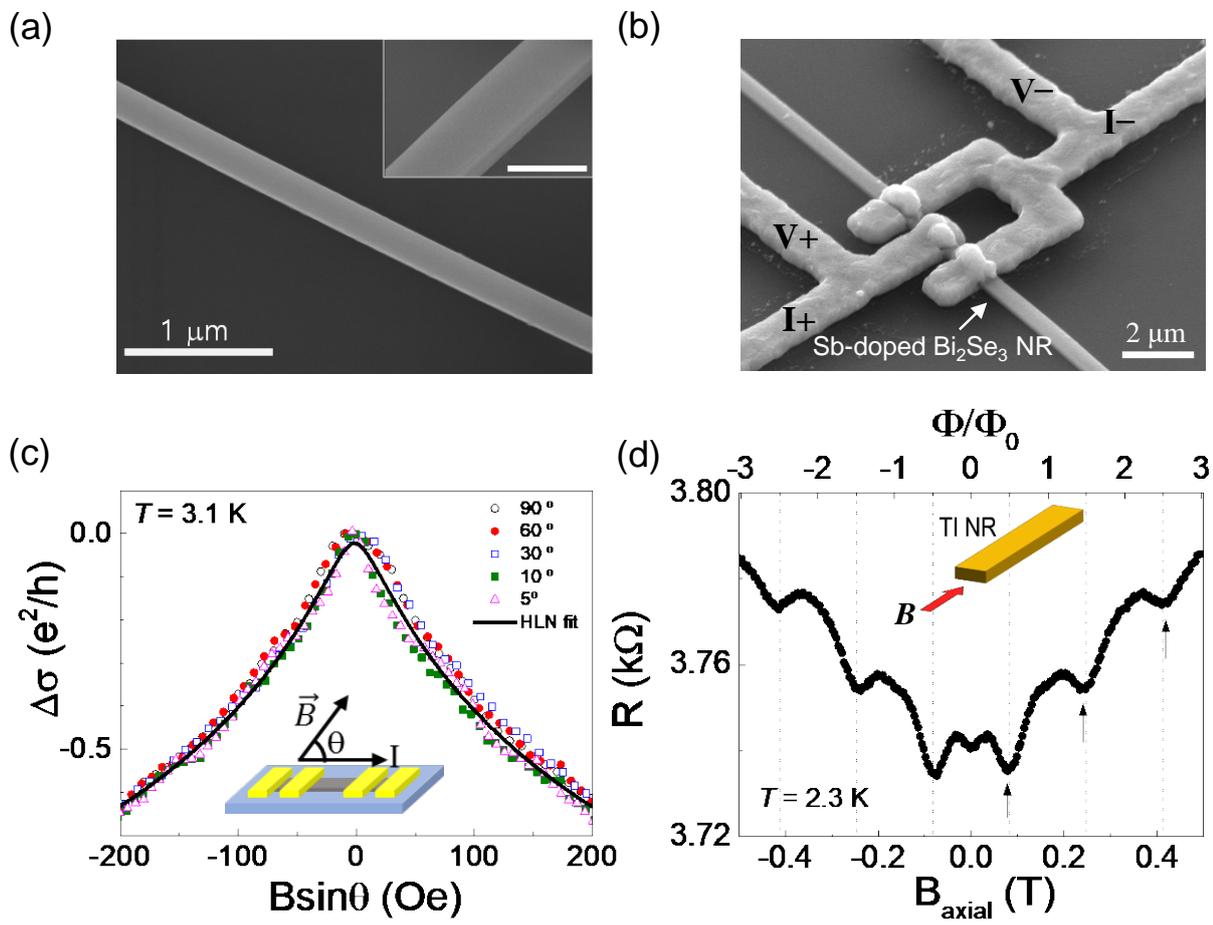

**Fig. 2**

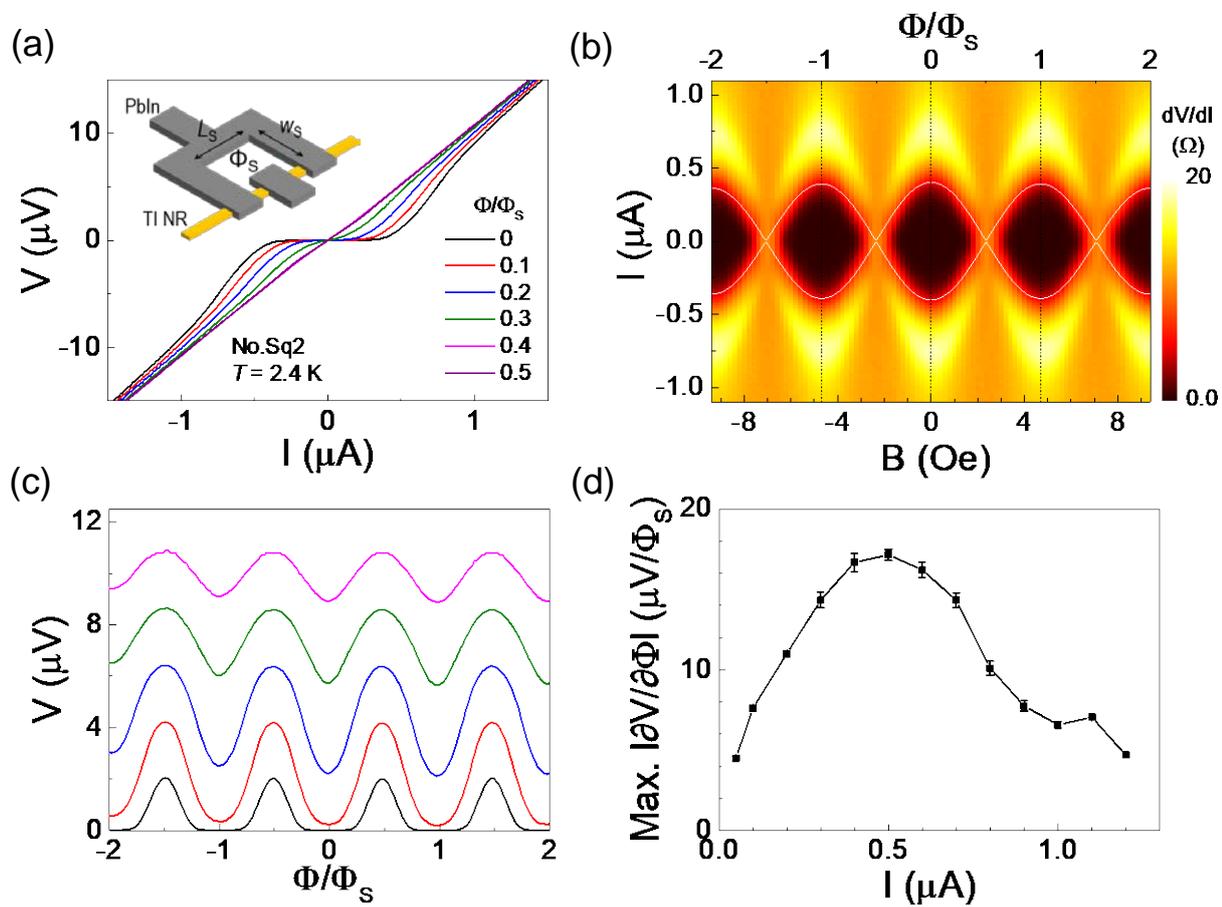

**Fig. 3**

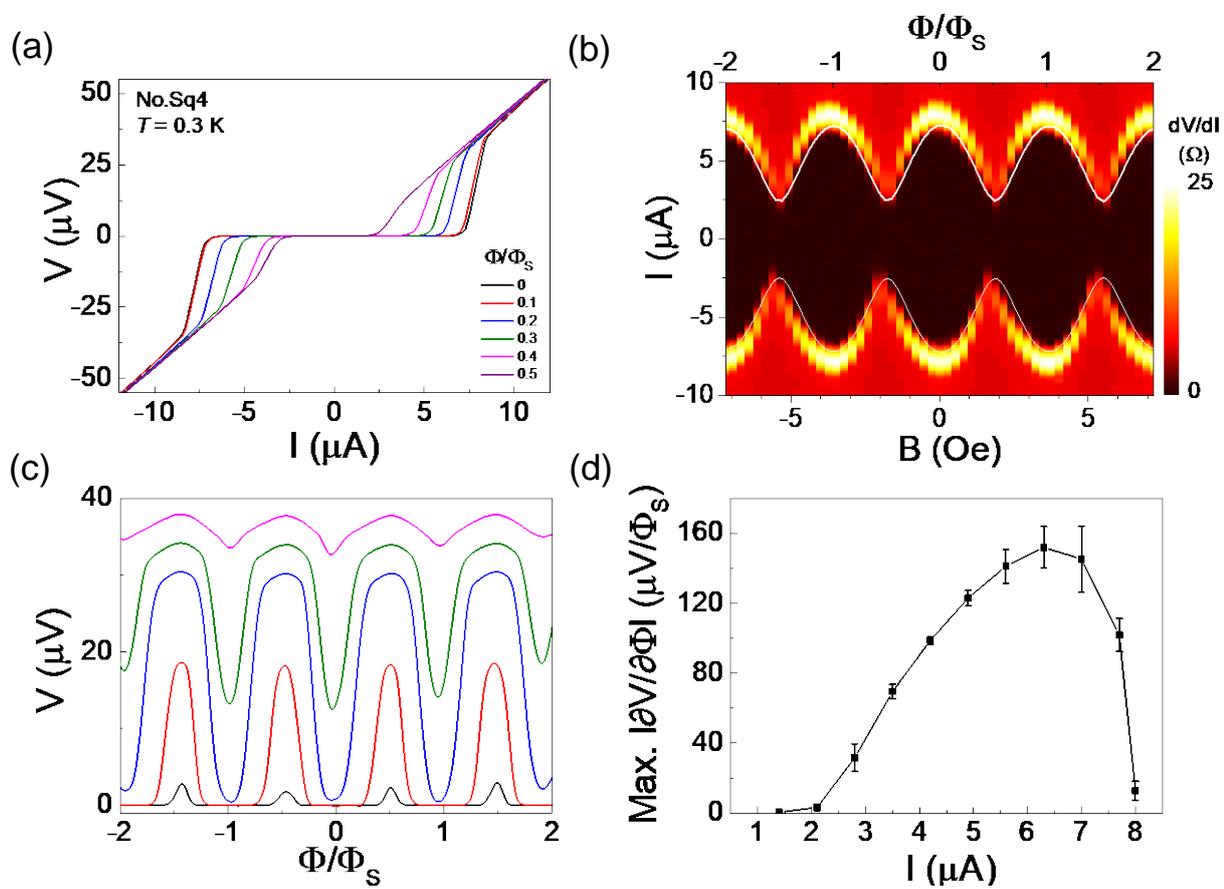

**Fig. 4**

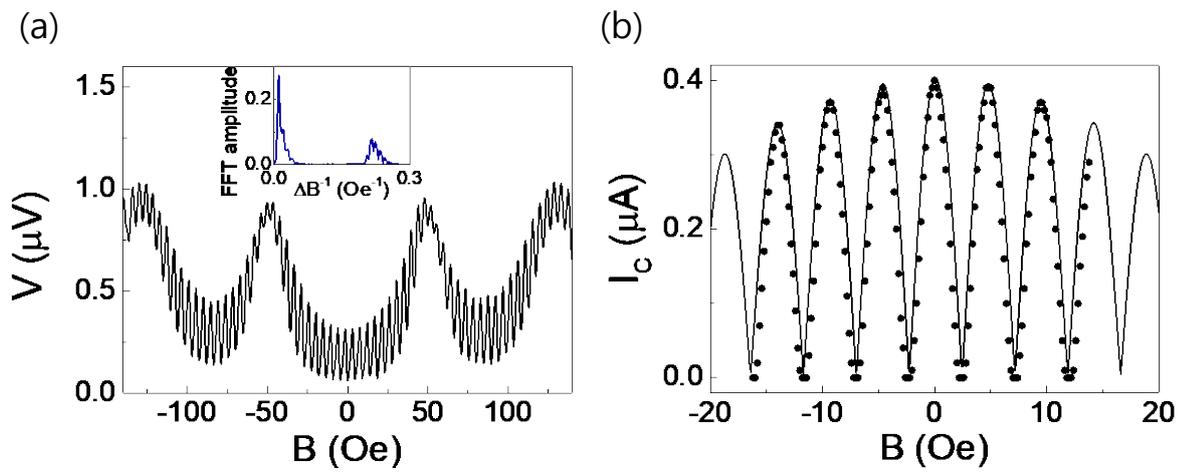